\documentclass[a4paper,11pt]{article}
\usepackage{pos}
\usepackage{hyperref}
\hypersetup{
    colorlinks=true,
    linkcolor=red,
    filecolor=magenta,      
    urlcolor=blue,
    citecolor = teal,
}

\usepackage{MnSymbol}

\newcommand{\bs}{\boldsymbol}
\DeclareMathOperator{\Tr}{Tr}

\title{Spatially inhomogeneous confinement-deconfinement phase transition in accelerated gluodynamics}
\ShortTitle{Spatially inhomogeneous confinement-deconfinement phase transition in accelerated \dots }

\author*[a]{Victor~V.~Braguta} 
\author[b,c]{Vladimir~A.~Goy} 
\author[a]{Jayanta~Dey}
\author[a]{Artem~A.~Roenko}

\emailAdd{vvbraguta@theor.jinr.ru}
\emailAdd{vovagoy@gmail.com}
\emailAdd{jayanta@theor.jinr.ru}
\emailAdd{roenko@theor.jinr.ru}

\affiliation[a]{Bogoliubov Laboratory of Theoretical Physics, Joint Institute for Nuclear Research,\\ 6  Joliot-Curie St., Dubna, 141980, Russia}
\affiliation[b]{Pacific Quantum Center, Far Eastern Federal University,\\
10 Ajax Bay, Russky Island,
Vladivostok, 690922, Russia}
\affiliation[c]{Institute of Automation and Control Processes, Far Eastern Branch, Russian Academy of Science,\\ 5 Radio Str., Vladivostok, 690041, Russia}

\abstract{This study explores confinement-deconfinement transition properties of SU($3$) Yang--Mills theory under weak accelerations at finite temperatures, using first-principles lattice simulations. The system is formulated in the Rindler spacetime, and the properties are studied from the perspective of a co-accelerating observer situated at the center of the lattice. 
We found that spatially separated confinement and deconfinement phases can coexist in the Rindler spacetime within certain intervals of temperature and acceleration. 
The position of the boundary between the phases is calculated as a function of temperature for several accelerations, and it is in accordance with the TE prediction, although a small deviation is observed. Moreover, in the weak acceleration regime, the critical temperature of the system is found to coincide with that of non-accelerated gluodynamics.}

\FullConference{The 42nd International Symposium on Lattice Field Theory (LATTICE2025)\\
2-8 November 2025\\
Tata Institute of Fundamental Research, Mumbai, India\\}

\begin{document}
\maketitle

\section{Introduction}\label{sec:intro}
``The effect of gravitation is locally indistinguishable from that of acceleration''--- the \textit{equivalence principle} from the general theory of relativity. This principle enables us to probe the influence of gravity on other fundamental forces by considering an accelerated reference frame. In particular, we are interested in the modification of the strong interaction by acceleration. This scenario is relevant for systems such as those in the vicinity of a black hole and for matter created in heavy-ion collisions (HICs). Studies~\cite{Kharzeev:2005iz, Prokhorov:2025vak} show that acceleration in HICs can reach values ranging from several hundred MeV to few GeV.  
Such large accelerations might lead to restoration of chiral symmetry in quantum chromodynamics (QCD).
Many theoretical studies (see \cite{Kou:2024dml} and references therein) have investigated the critical acceleration associated with the restoration of chiral symmetry. In contrast, the confinement-deconfinement transition under acceleration remains largely unexplored. In the present study, we address this aspect in the regime of weak acceleration at finite temperature. 

The condition for thermodynamics equilibrium in gravity is given by the Tolman-Ehrenfest (TE) law~\cite{Tolman:1930ona}. According to this law, the local equilibrium temperature of a system under the influence of static gravitational field is coordinate dependent and is given by  $T(\bs r) = \textit{const}/\sqrt{g_{00}(\bs r)}$, where $g_{00}$ is the time-time component of the metric.
The first non-perturbative study for SU(3) Yang-Mills theory under weak acceleration was reported recently in Ref.~\cite{Chernodub:2024wis}.
There, the authors assumed the validity of the TE law and incorporated acceleration via the corresponding temperature gradient.
The present study is based on another approach.
Here, we formulate the system in the Rindler coordinate and employ a first-principles non-perturbative technique---lattice simulation in this curved background.   
We find that the results for accelerated gluodynamics are consistent with the TE law, with only small deviations. 
In the Rindler spacetime, the metric exhibits a singularity that causally separates the left and right Rindler wedge. As a result, a co-accelerating observer perceives Minkowski vacuum as heat bath at the Unruh temperature. 
Here, we study the system in the right Rindler wedge, and the considered range of acceleration $\alpha$ is very small compare to the characteristic energy scale of gluodynamics, i.e., $\alpha \ll \Lambda \approx 200$~MeV. Consequently, the effects of the Unruh temperature are negligible relative to the systematic uncertainties and thus not measurable.


\section{Gluodynamics in Rindler spacetime}\label{sec:Theory_Rindler} \label{sec:form}
The comoving frame for an accelerated observer can be described by the Rindler coordinates~\cite{Crispino:2007eb, Braguta:2026nfy}.
In the 3+1 dimensions, with acceleration $\alpha$ directed along the $z$-axis, the invariant interval in the Rindler spacetime takes the form
\begin{equation}
ds^2 = g_{\mu \nu} dx^{\mu} d x^{\nu} =(1+{\alpha} z)^2 dt^2 - dx^2 - dy^2 -dz^2 \,,
\label{eq:rindler}
\end{equation}
where $dx^{\mu}=(dt, dx, dy, dz)$. Here, the metric has a singularity at $z_h=-{1}/{\alpha}$, called the Rindler horizon.
 
To compute the partition function using the path integral formalism, it is customary to work in imaginary time, $\tau = i t$. The interval~(\ref{eq:rindler}) in this case transforms to
\begin{equation}
ds^2 = (g_E)_{\mu \nu} dx_E^{\mu} dx_E^{\nu}=- (1+{\alpha} z)^2 d\tau^2 - dx^2 - dy^2 -dz^2\,,
\label{eq:Rindler_metric}
\end{equation}
where $dx_E^{\mu}=(dx, dy, dz, d \tau)$. Hereafter, the subscript $E$ stands for Euclidean space.
The partition function for gluodynamics can be written as the integral over gluon degrees of freedom 
\begin{equation}
\mathcal{Z} = \int \! D A \exp {\left ( -S_E \right )}\,.
\end{equation}
Here, $S_E$ is the Euclidean gluon action, which in the Rindler space takes the form:
\begin{equation}
 S_E = \frac 1 {4 g_{YM}^2} \int \! d^4 x \sqrt g_E g_E^{\mu \nu} g_E^{\alpha \beta} F^a_{\mu \alpha} F^a_{\nu \beta} =  \frac 1 {4 g_{YM}^2} \int d^4 x \left ( \frac 1  {1 +{\alpha} z} {\left({\bs E}^{a} \right)^2} + (1+{\alpha} z) \left({\bs H}^{a} \right)^2
 \right ) \,, 
 \label{eq:SE_continuum}
\end{equation}
where $g_{YM}$ is the strong coupling constant, $F_{\mu \nu}^a$ denotes the gluon field strength tensor, $\bs E^a$ and $\bs H^a$ are chromoelectric and chromomagnetic gluon fields, respectively, and $(g_E)_{\mu\nu}$ is the Rindler metric in Euclidean space.

In the imaginary time formalism, $\tau$ is a compact variable with period $T_0$ i.e., $\tau \in (0, 1/T_0)$. In Rindler spacetime, the time interval scales as $d\tau \to (1+\alpha z) d\tau$. Hence, the local temperature can be expressed as
\begin{equation}\label{eq:TE}
T(z) = \frac{T_0}{1+ \alpha z}\, ,
\end{equation}
which is the Tolman-Ehrenfest law in the Rindler spacetime. 
Note that the local acceleration is also non-uniform and depends on the $z$-coordinate~\cite{Braguta:2026nfy}
\begin{equation}\label{eq:acceleration_z}
    \alpha(z) = \frac{\alpha_0}{1+\alpha z}\, .
\end{equation}
At the observer position, $z=0$, acceleration and temperature are $\alpha \equiv \alpha_0$ and $T\equiv T_0$, respectively.

\section{Lattice setup}\label{sec:Theory_setup}
To discretize the continuum action (\ref{eq:SE_continuum}) on lattice, we employ the tree-level improved Symanzik action~\cite{Curci:1983an, Luscher:1985zq}.  To incorporate the Rindler metric~\eqref{eq:rindler} into the gluon action, we multiply the chromoelectric and chromomagtetic lattice plaquettes/rectangles by the factors $1/(1+\alpha_l z)$ and  $(1+\alpha_l z)$, respectively. Thus, the lattice discretizations of the continuum action~(\ref{eq:SE_continuum}) reads 
\begin{multline}\label{eq:SE_lattice}
S_{E} = \beta  \sum_{x}\bigg[
\frac{1}{1 + \alpha_l z} \sum_{i} \bigg(c_0  \left(1 - \frac{1}{N_c} \text{Re} \Tr \bar{U}_{0i}\right) + 
c_1 \left(2 - \frac{1}{N_c} \text{Re} \Tr \big(\bar{W}_{0i}^{1\times2} + \bar{W}_{i0}^{1\times2}\big) \right) \bigg) + {}
 \\
{} +
(1 + \alpha_l z) \sum_{j>k} \bigg(c_0  \left(1 - \frac{1}{N_c} \text{Re} \Tr \bar{U}_{jk}\right) + 
c_1 \left(2 - \frac{1}{N_c} \text{Re} \Tr \big(\bar{W}_{jk}^{1\times2} + \bar{W}_{kj}^{1\times2}\big) \right) \bigg)
\bigg]\, ,
\end{multline}
where $\beta = 6/g_{YM}^2$ is an inverse lattice gauge coupling,  $\alpha_l=\alpha a$ is the acceleration in lattice units, $\bar{U}_{\mu\nu}$ denotes the clover-type average of four plaquettes, $\bar{W}_{\mu\nu}^{1\times2}$ is the clover-type average of four rectangular loops, and $c_0 = 1 - 8 c_1$, $c_1 = -1/12$. 
Since the lattice action~\eqref{eq:SE_lattice} is free of the sign problem, standard lattice techniques can be directly applied to study gluodynamics in Rindler spacetime with real acceleration. 

For simulations, we use a \textit{base lattice} of size $N_t \times N_x \times N_y \times N_z = N_t \times N_s^2 \times N_z \equiv 5\times 40^2 \times  121$. For the continuum-limit extrapolation, three more temporal extent $N_t = 4, 6, 8$ are used, keeping the aspect ratio fixed as $N_s/N_t=8$ and $(N_z-1)/N_t=24$. Moreover, to estimate the finite volume effects, in addition to the \textit{base lattice}, we use $N_s=50, 60$ for transverse plane and $N_z=151, 181$ for longitudinal extension. 
Unless otherwise specified, the results discussed in subsequent sections pertain to our \textit{base lattice}.
For scale setting $a(\beta)$, we use the string tension  from Ref.~\cite{Beinlich:1997ia}, taking $\sqrt{\sigma} = 440$~MeV.
For reference, we use the critical temperature $T_{c0}$ of the standard gluodynamics obtained from the infinite-volume extrapolation results for the critical coupling $\beta_c$ from Refs.~\cite{Beinlich:1997ia, Borsanyi:2022xml}.

The action~(\ref{eq:SE_lattice}) assumes that the observer is located at $z=0$. We adjust the geometry of our lattices  so that the origin of the coordinate system coincides with the center of the lattice in the $z$-direction. Consequently, in physical units, the $z$-coordinate varies over the interval  $z \in [-(N_z - 1)a/2, (N_z - 1)a/2]$.\footnote{In our simulations $N_z$ is an odd number.} 
The main part of calculations are done with open boundary conditions implemented along this direction. To ensure that the results are not affected by the boundaries for used lattice sizes, we also repeat simulation on the \textit{base lattice} with Dirichlet boundary conditions and found excellent consistency in the results.
For the remaining lattice directions---$x,y,\tau$---we use the standard coordinates and impose periodic boundary conditions.

We perform simulations for accelerations $\alpha = 3, 6, 9, 18$~MeV. Thus, the condition {$\alpha\ll \Lambda$} holds throughout our analysis. Moreover, the lattice parameters are adjusted so that the entire lattice volume remains well away from the Rindler horizon at $z_h = -1/\alpha$, and large accelerations $\alpha(z) \sim \Lambda$ (according to Eq.~(\ref{eq:acceleration_z})) are never reached in studied volume. 

\section{The observables}\label{sec:observables}
In standard gluodynamics, the Polyakov loop serves as the order parameter for the confinement-deconfinement phase transition. It is also known as the thermal Wilson loop, defined via the path-ordered integral of the temporal gauge field as
\begin{align} \label{eq:L_continuum}
    L({\bs r}) = \Tr\, \mathcal{P} \exp \biggl(\oint_{0}^{1/T} d \tau A_4( {\tau},{\bs r}) \biggr) \,,
\end{align}
where ${\mathcal P}$ is the path-ordering operator. In confinement phase, the Polyakov loop respects the discrete center $\mathbb{Z}_3$-symmetry, so that $\langle L \rangle = 0$. At finite temperatures, this symmetry is spontaneously broken and system enters the deconfinement phase, where $\langle L \rangle \neq 0$. In the thermodynamic limit, for pure SU($3$) Yang-Mills theory, this transition is of the first-order. 

On the lattice, the continuum expression~(\ref{eq:L_continuum}) is discretized as
\begin{equation}\label{eq:L_lattice_bare}
    L^b(\bs r) = \frac{1}{N_c} \Tr \left[ \prod_{\tau = 0}^{N_t - 1} U_4(\tau, \bs r) \right]\,,
\end{equation}
where $U_4(\tau, r)$ denotes the temporal component of the lattice gauge field. Here, $L^b(\bs r)$ is the bare Polyakov loop, which contains ultraviolet divergences.
In the standard gluodynamics, the Polyakov loop can be renormalized using the multiplicative factor $Z(g^2)$, calculated in Refs.~\cite{Kaczmarek:2002mc,Gupta:2007ax}.
In the Rindler spacetime, the scheme of multiplicative renormalization modifies (see details in Ref.~\cite{Braguta:2026nfy}), and we compute the renormalized local Polyakov loop as follows
\begin{equation}\label{eq:L_z_lattice}
    L(z) = \left(Z(g^2)\right)^{N_t (1+\alpha z)} \cdot \frac{1}{N_s^2 \delta z} \sum_{\bar z} \sum_{x,y} L^b(x,y,\bar z)\,,
\end{equation}
where $L^b(x, y, z) = L^b(\bs r)$.
Here, we average over a three-dimensional volume that extends over the full lattice in the $x$ and $y$ direction and over a small thickness $\delta z$ along the $z$-direction. In physical units, this thickness corresponds to $\delta z_{\rm ph} =  \delta z \cdot a$. The coordinate $z$ is taken as the center of $\delta z$ interval. Unless specified, results discussed in the subsequent sections are for $\delta z = 1$. 
For the analysis of transition,  we fit the Polyakov loop using the function
\begin{equation}\label{eq:fit_tanh}
    L(z) = C_0 \tanh\left( \frac{z-z_c}{\Gamma}\right) + C_1\,,
\end{equation}
where $C_0, C_1, \Gamma$, and $z_c$ are the fit parameters. 

For the analysis of confinement-deconfinement transition, we also use the renormalized Polyakov loop susceptibility, $\chi(z)$, which account for fluctuations in the Polyakov loop. Here, we compute it using the formula
\begin{gather}\label{eq:chi_z_ren}
T^3 \chi(z) =  \frac{N_s^2\, \delta z}{N_t^3} \left( \langle |L(z)|^2 \rangle  - \langle |L(z)|\rangle^2 \right)\,,
\end{gather}
where the temperature in lattice is calculated as $T=1/(a N_t)$.
For analysis, the susceptibility data are fitted by the Gaussian function
\begin{equation}\label{eq:fit_Gauss}
    \chi(z) = A_0 \exp\left(-\frac{(z-z_c)^2}{2 \sigma^2}\right) + A_1 + A_2 (z-z_c)\,,
\end{equation}
where $A_0, A_1, A_2, z_c, \sigma$ are the fit parameters.
The parameter $z_c$ in Eq.~(\ref{eq:fit_Gauss}) determine the position of the susceptibility peak, and that from Eq.~(\ref{eq:fit_tanh}) determines the position of inflection point of the Polyakov loop. 
Below we compare the results for two definitions of the transition coordinate, $z_c$.


\section{Results}\label{sec:res}

\begin{figure}[t]
\centering
\includegraphics[width=.49\textwidth]{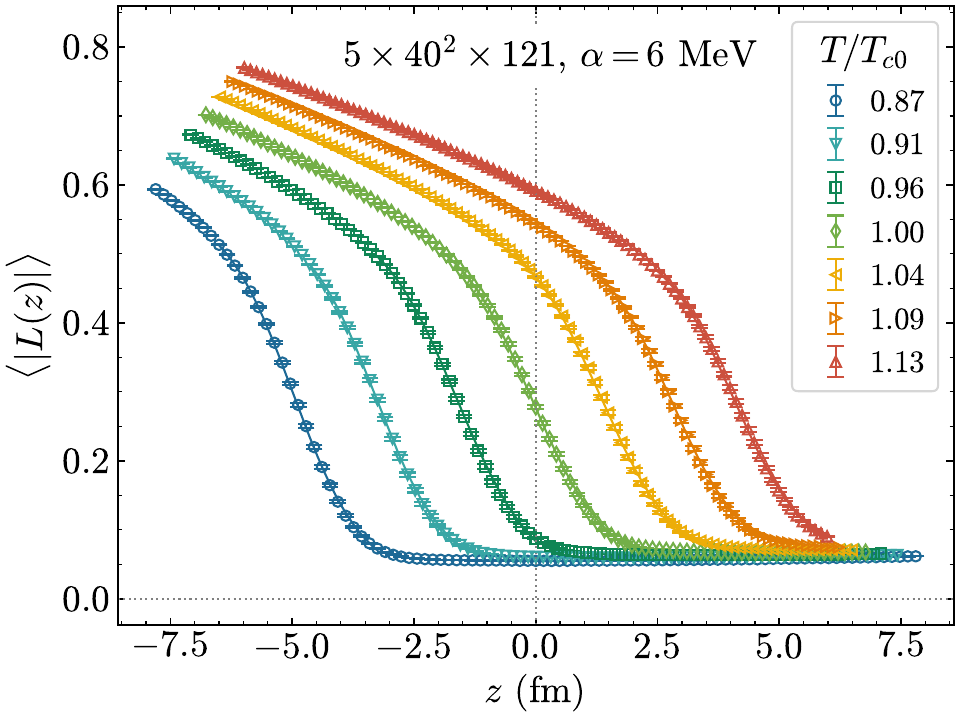}
\hfill
\includegraphics[width=.49\textwidth]{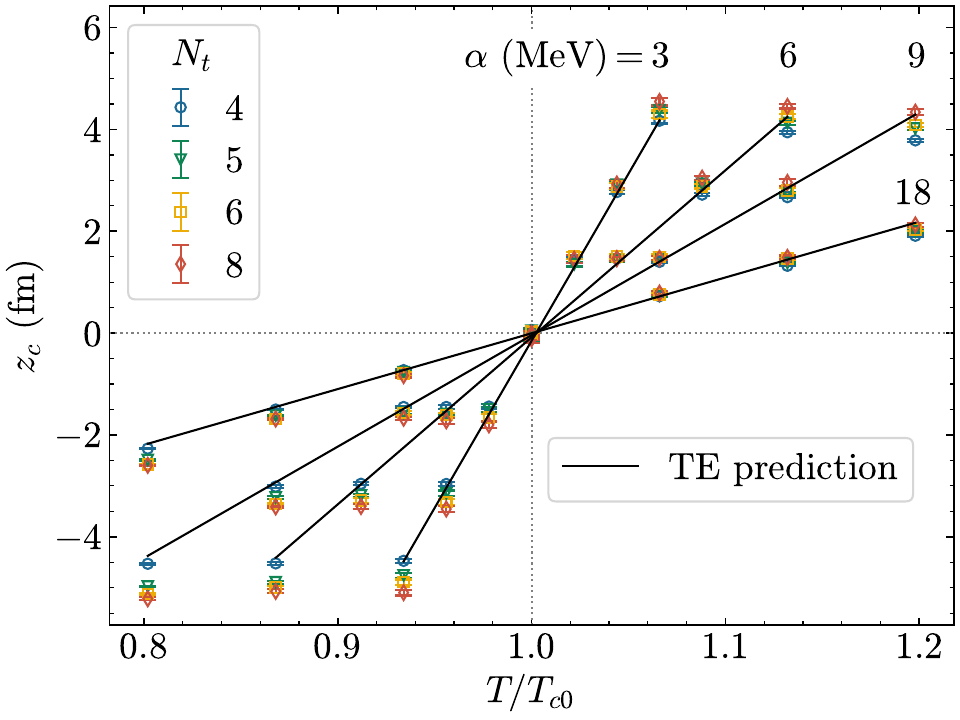}
\caption{(Left) Renormalized local Polyakov loop as a function of coordinate $z$.
(Right) The critical distances as a function of temperature.}
\label{fig:1}
\end{figure}

The left panel of Fig.~\ref{fig:1} shows the renormalized local Polyakov loop as a function of coordinate $z$ for our \textit{base lattice} at a fixed acceleration $\alpha=6$~MeV for various temperature in the vicinity of critical temperature $T_{c0}$ of the corresponding non-accelerated system. Here, we use modulus of the Polyakov loop, $\langle|L(z)|\rangle$, which remains small but nonzero even in the confined phase at finite volume and vanishes only in the infinite-volume limit. From the figure, we can see spatial coexistence of confined and deconfined phase along the $z$-coordinate---confined in $z\gtrsim 0$ and deconfined in $z\lesssim 0$. The position of confinement-deconfinement transition depends on temperature and shifts in the direction of acceleration as the temperature increases. To find the position of transition, which we called the critical distance $z_c$, we use two approaches, discussed in Section~\ref{sec:observables}.
First, we will present the results from the Polyakov loop inflection point.

We determine $z_c$ by fitting the Polyakov loop using Eq.~(\ref{eq:fit_tanh}). The right panel of Fig.~\ref{fig:1} shows $z_c$ as a function of temperature $T/T_{c0}$ for the lattices with $N_t = 4,5,6,8$ at several accelerations. 
These data are compared with  the prediction $z_c^{\rm TE}$ based on the Tolman-Ehrenfest law.
Assuming that the local temperature at $z_c$ is equal to the critical temperature $T_{c0}$ of standard gluodynamics, we obtain from Eq.~(\ref{eq:TE})
\begin{equation}\label{Eq:zc_TE}
    z_c^{\rm TE} = \frac{1}{\alpha} \left(\frac{T}{T_{c0}} - 1 \right)\, .
\end{equation}
The figure shows that, apart from small deviations, all data points from our simulation are well described by the TE prediction (black solid lines).

\begin{figure}[t]
\centering
\includegraphics[width=.49\textwidth]{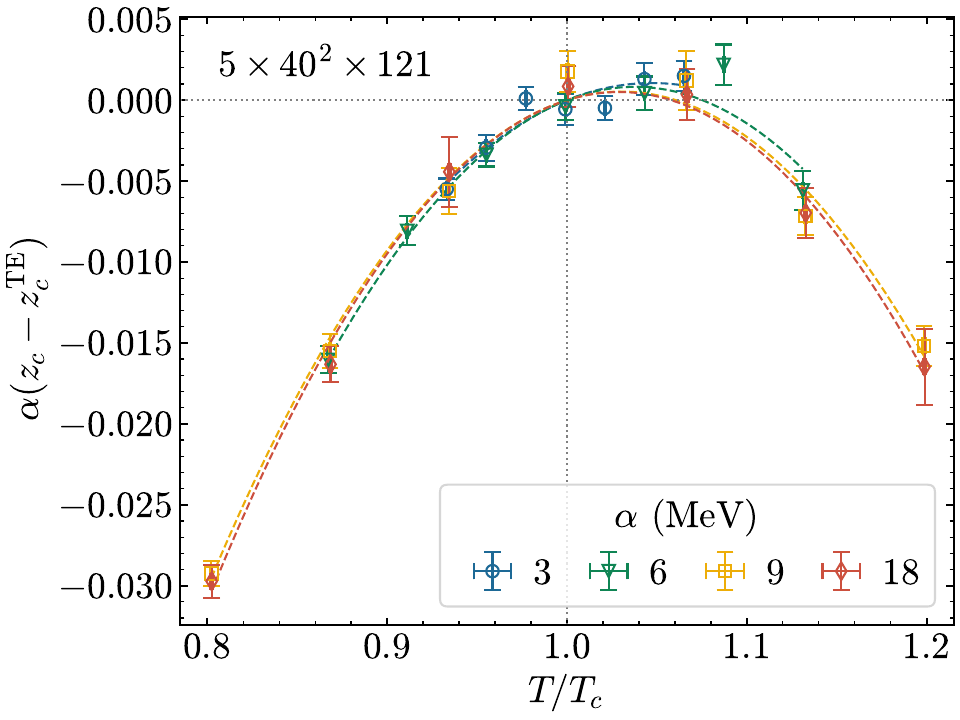}
\hfill
\includegraphics[width=.49\textwidth]{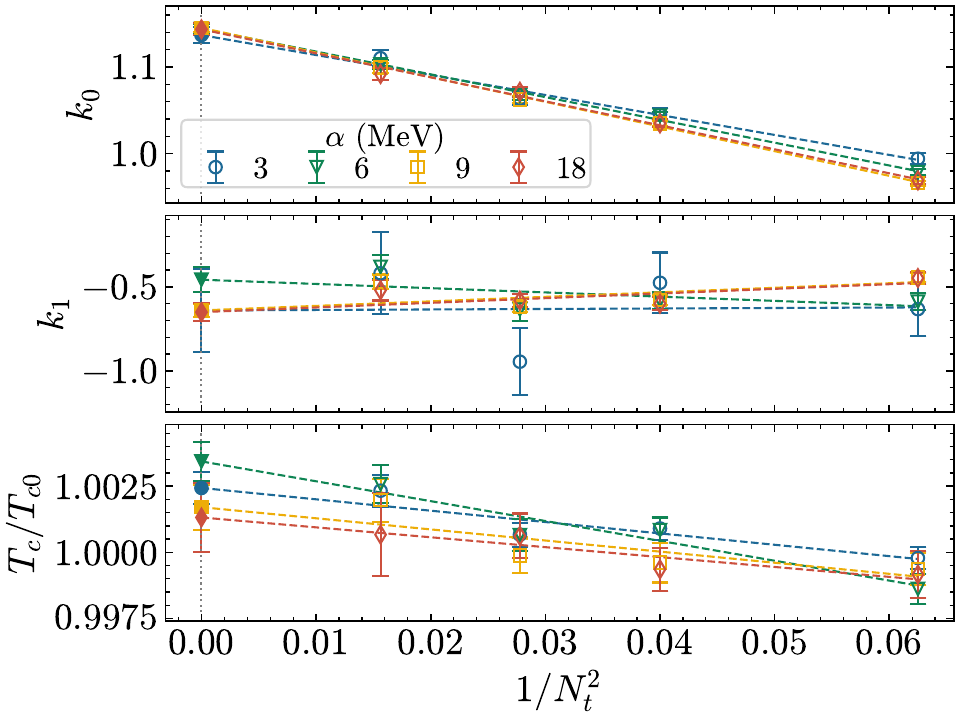}
\caption{(Left)  Difference between $z_c$ and $z_c^{\rm TE}$ as a function of temperature. (Right) 
Continuum limit extrapolation of the fit parameters $k_0, k_1$, and $T_c$ for the critical distance obtained from the inflection point of the local Polyakov loop.}
\label{fig:2}
\end{figure}

To quantify the deviation of the results from the TE prediction, we fit the data with the function
\begin{equation}
\label{eq:zc_fit}
    z_c = \frac{k_0}{\alpha} \left(\frac{T}{T_{c}}  - 1\right) + \frac{k_1}{\alpha} \left(\frac{T}{T_{c}}  - 1\right)^2\,,
\end{equation}
where $k_0$, $k_1$, and $T_c$ are the fit parameters. Notice that the fit parameter $T_c$ is different form critical temperature $T_{c0}$ of non-accelerated system. In accelerated gluodynamics, the critical temperature may depend on acceleration, and $T_c$ allow us to determine this dependence. However, for weak accelerations we expect $T_c (\alpha) \approx T_{c0}$. We will verify this shortly. 

The left panel of Fig.~\ref{fig:2} shows the difference between the critical distance $z_c$ obtained from lattice simulation and the TE prediction, plotted as a function of temperature scaled by the fit parameter $T_c$.
Here we use $z_c^{\rm TE} = (T/T_c - 1)/\alpha$, where a possible change in $T_c$ by acceleration is taken into account.
Horizontal axis includes uncertainty propagated from $T_c$. Scaling the vertical axis by acceleration shows that the deviation is largely independent on acceleration. We observe a parabolic deviation that increases as we move away from $T_c$.

The fit function~(\ref{eq:zc_fit}) describes the data very well. The resulting best-fit parameters are plotted as function of $1/N_t^2$ in the right panel of Fig.~\ref{fig:2}. Open markers represent the data, which are fitted by a linear function of the form $\mathcal{O} = \mathcal{O}_0 + \mathcal{O}_1 / N_t^2$, shown as dashed lines. Continuum limit is obtained by extrapolating to $N_t \rightarrow \infty$, and the results at this limit are shown by the filled marker at $1/N_t^2 = 0$. 
For an exact match with the TE prediction, the value of the fit parameters should be, within uncertainties, $k^{\rm TE}_0=1, \, k^{\rm TE}_1 = 0$.
However, in the continuum limit we find that the leading order coefficient $k_0$  deviates from this value by more than $10\%$, and $k_1$ is also non-zero. The fit parameter $T_c$ differs from $T_{c0}$ by less than $0.3\%$, which allow us to conclude that, for weak acceleration, critical temperature of accelerated gluodynamics remain the same as that of non-accelerated system.

\begin{figure}[t]
\centering
\includegraphics[width=.49\textwidth]{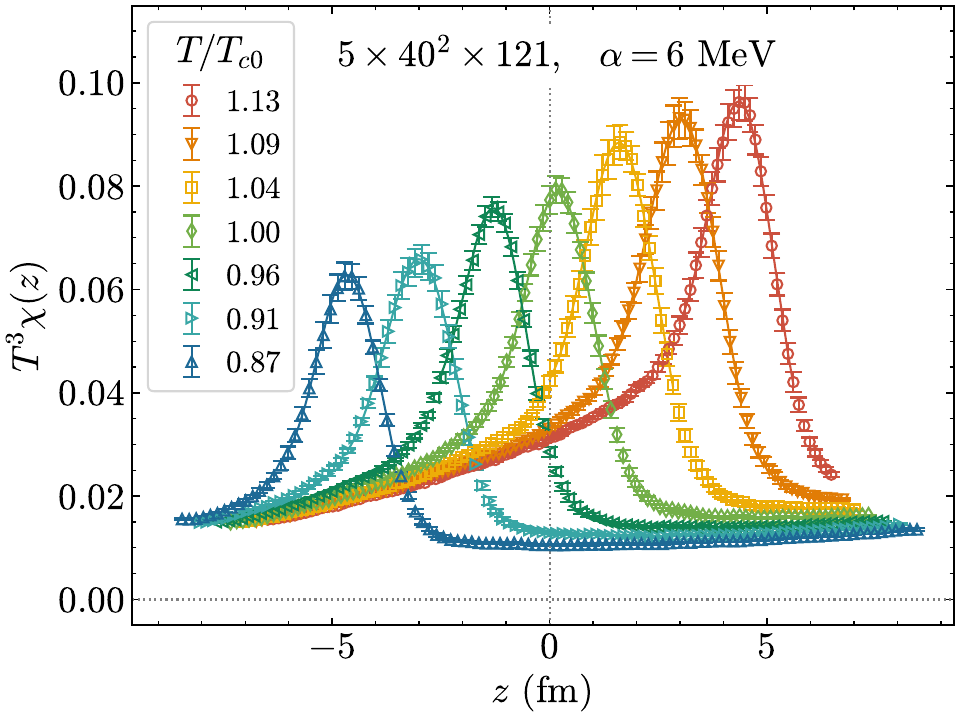}
\hfill
\includegraphics[width=.49\textwidth]{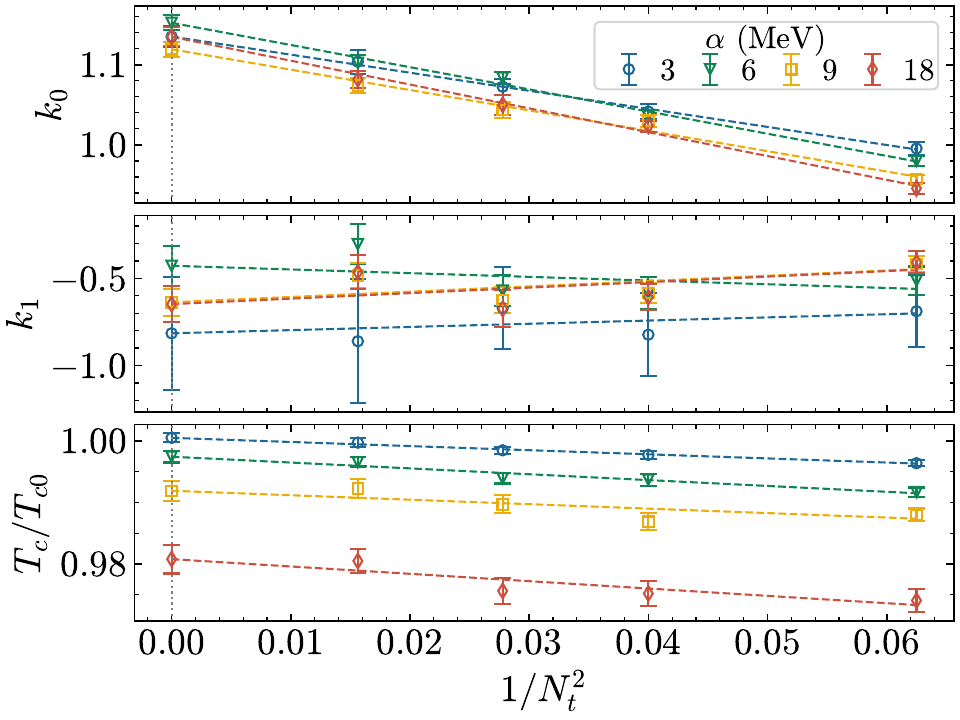}
\caption{(Left) Renormalized Polyakov loop susceptibility as a function of the coordinate $z$. (Right) Continuum limit extrapolation of the fit parameters $k_0, k_1$, and $T_c$ for the critical distance obtained from the peak position of the local susceptibility.}
\label{fig:3}
\end{figure}

In lattice gauge theory, the Polyakov loop susceptibility, Eq.~(\ref{eq:chi_z_ren}), is also commonly used to study the properties of the confinement-deconfinement transition. In the vicinity of the transition, fluctuations in the Polyakov loop become pronounced, and the susceptibility exhibits a peak-like structure. The left panel of Fig.~\ref{fig:3} shows this susceptibility as a function of the coordinate $z$ for various temperatures at acceleration $\alpha=6$~MeV for our \textit{base lattice}. To determine the the critical distance, we fit the susceptibility peak using the Gaussian function given in Eq.~(\ref{eq:fit_Gauss}). As in the case of the Polyakov loop inflection point discussed above, we study the deviation of the results from the TE prediction.

The right panel of Fig.~\ref{fig:3} is analogous to the right panel of Fig.~\ref{fig:2}, but obtained from the susceptibility peaks. In the continuum limit, $k_0$ and $k_1$ are largely the same as those extracted from the inflection point of the Polyakov loop. The parameter $T_c$ exhibits a mild dependence on acceleration. However, we found that the deviation in $T_c$ from $T_{c0}$ is related to the thickness $\delta z$ of the transverse plane over which the Polyakov loop is averaged (see Eq.~(\ref{eq:L_z_lattice})).     


\begin{figure}[t]
\centering
\includegraphics[width=.49\textwidth]{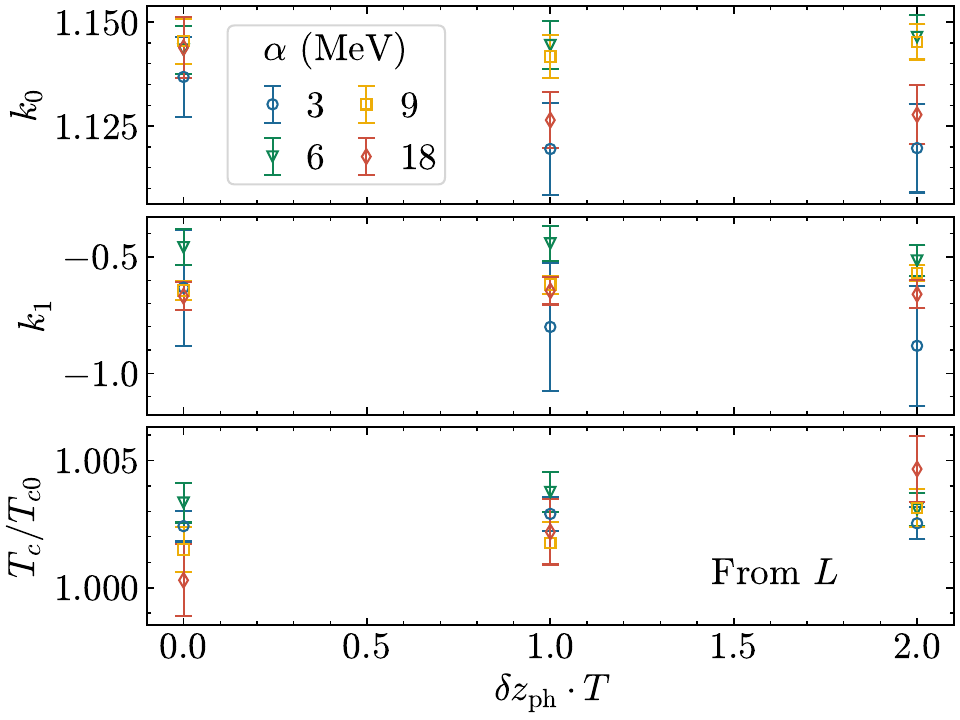}
\hfill
\includegraphics[width=.49\textwidth]{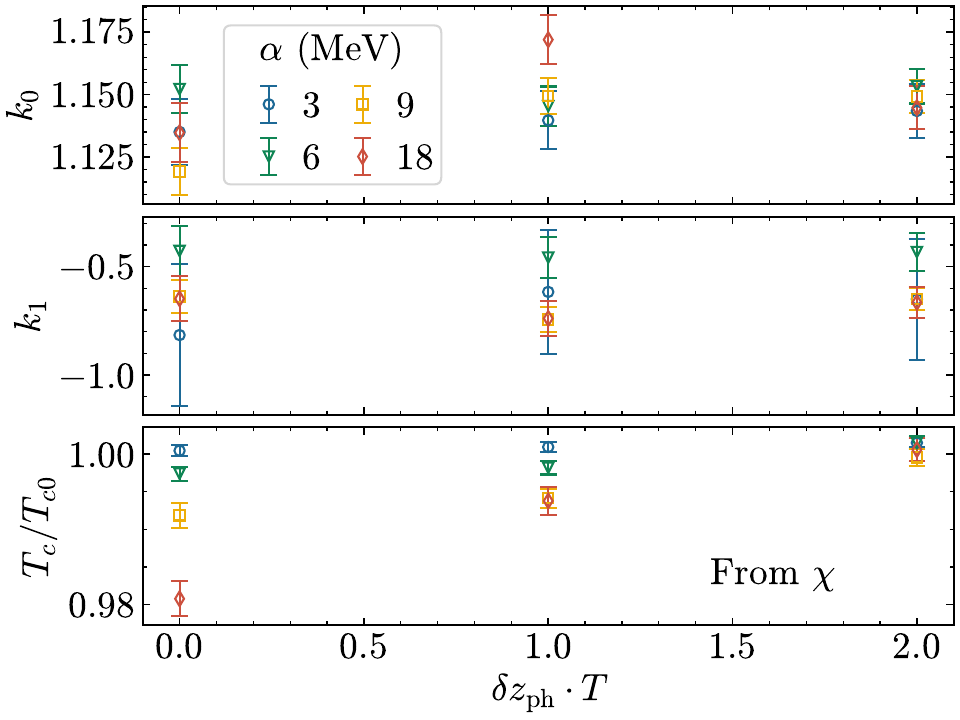}
\caption{Continuum limit values of the fit parameters $k_0, k_1, T_c$ obtained from the Polyakov loop inflection point (left) and susceptibility peak (right) as a function the thickness $\delta z$. }
\label{fig:4}
\end{figure}

To compare the results obtained from the two definitions of the critical distance---the Polyakov loop inflection and the susceptibility peak---we plot the continuum limit values of the fit parameters $k_0,\, k_1,\, T_c$ as functions of thickness $\delta z = 1, N_t, 2N_t$ (corresponds to $\delta z_{\rm ph} \cdot T = 0, 1, 2$ in continuum limit) in the left and right panels of Fig.~\ref{fig:4}, respectively.
From the left panel, we see that the results are largely stable for thickness up to $\delta z = 2 N_t$.
From the right panel, we observe that only the parameter $T_c$ exhibits a weak dependence on thickness. For $\delta z = 1$, $T_c$ deviates from $T_{c0}$ for larger accelerations. However,  this deviation diminishes as $\delta z$ increases, and for $\delta z=2 N_t$ we find $T_c(\alpha) \approx T_{c0}$. Notice that for $\delta z=1$, the physical thickness in the continuum limit is $\delta z_{\rm ph} = 0$, so the observables---the Polyakov loop and its susceptibility---correspond to purely two-dimensional averaging. In contrast, for $\delta z =N_t$ or $2N_t$, the averaging remains three dimensional even in the continuum limit. Hence, for three dimensional averaging, the smallest thickness is $\delta z=N_t$, where both definitions yield consistent results. 

To estimate the finite volume effects associated with the longitudinal ($N_z$) and transverse ($N_s$) lattice sizes, we simulate the system on the lattices $5\times 40^2 \times N_z$ with $N_z = 121, 151, 181$ and $5\times N_s^2 \times 121$ with $N_s = 40, 50, 60$.
We found that, in the limits $N_z \to \infty$ and $N_s \to \infty$, the parameter $k_0$ increases by approximately $1\%$, $k_1$ remain largely independent, within uncertainty, and the critical temperature $T_c$ changes less than $0.3\%$. Therefore, the small deviation from the TE law persists in the large volume limit. We also checked that it is not an artifact of the open boundary conditions by simulating the \textit{base lattice} with Dirichlet boundary conditions in the $z$-direction. Moreover, we studied the nature of the confinement-deconfinement transition using the height and width of the susceptibility peak in the infinite volume limit.%
\footnote{Note, however, that one cannot extend the system to infinity in the $z<0$ direction, as the horizon, $z_h = -1/\alpha$ should lie outside the studied volume.}
We found that the spatial transition is of crossover type (as also observed in Ref.~\cite{Chernodub:2024wis}), with the minimum width restricted by the corelation length of the system. Further details can be found in Ref.~\cite{Braguta:2026nfy}.

\section{Summary}\label{sec:summary}

In this paper, we studied the influence of weak acceleration on the phase diagram of SU($3$) Yang--Mills theory.
To incorporate acceleration, we formulated the system in the Rindler spacetime, and carried out the study using first-principles non-perturbative technique of lattice simulation. 
This approach differs from the formulation of accelerated theory based on the local temperature gradient, as it was incorporated in Refs.~\cite{Chernodub:2024wis}, and may be used for large accelerations.

The Rindler metric exhibits a singularity, known as the Rindler horizon. In our study, the entire lattice volume was kept far from this horizon within the right Rindler wedge.
To compute the gluon action on lattice, we used the tree-level improved Symanzik action. The Rindler metric is incorporated by multiplying the chromoelectric and chromomagnetic parts of the lattice action by the corresponding spatially dependent coefficients.

We computed the renormalized Polyakov loop and its susceptibility to investigate the phase structure of accelerated gluodynamics and observe a spatial confinement-deconfinement crossover transition in the system. The resulting spatial configuration---confinement in the direction of acceleration and deconfinement in the opposite direction---is in accordance with the TE law.
The position of transition---critical distance---is also well described by the TE law, with a small deviation of approximately $10\%$ in the leading-order coefficient. Within the studied acceleration regime and up to very good accuracy, the critical temperature in accelerated system remain unchanged from that of standard homogeneous system.  

It is interesting to compare this results with recent lattice studies of the phase diagram for rotating gluodynamics~\cite{Braguta:2021jgn,Braguta:2023iyx,Braguta:2024zpi}.
There, the rotating system is studied using the similar approach: it is simulated in corotating reference frame where external gravitational field appears.
At certain values of parameters, the spatial transition occurs in rotating system, which separates the central deconfined region from confined one at larger radial distance~\cite{Braguta:2023iyx, Braguta:2024zpi}.
This configuration contradicts with the expectations from TE law based on local equilibrium temperature due to specific asymmetry between chromomagnetic and chromoelectric fields in the action, induced by gravity~\cite{Braguta:2024zpi}.
For the Rindler metric, the asymmetry between the components of gluon fields also appears, but it has more simple form, because the action is invariant with respect to the observer's position (see details in Ref.~\cite{Braguta:2026nfy}).

\acknowledgments
This work has been carried out using computing resources of the Federal collective usage center Complex for Simulation and Data Processing for Mega-science Facilities at NRC ``Kurchatov Institute'', http://ckp.nrcki.ru/, and
the heterogeneous computing platform HybriLIT (LIT, JINR).
This work was supported by the Russian Science Foundation (project no. 23-12-00072).

\bibliographystyle{JHEP}
\bibliography{gravity.bib}

\providecommand{\href}[2]{#2}\begingroup\raggedright\begin{thebibliography}{10}

\bibitem{Kharzeev:2005iz}
D.~Kharzeev and K.~Tuchin, \emph{{From color glass condensate to quark gluon plasma through the event horizon}}, \href{https://doi.org/10.1016/j.nuclphysa.2005.03.001}{\emph{Nucl. Phys. A} {\bfseries 753} (2005) 316} [\href{https://arxiv.org/abs/hep-ph/0501234}{{\ttfamily hep-ph/0501234}}].

\bibitem{Prokhorov:2025vak}
G.Y.~Prokhorov, D.A.~Shohonov, O.V.~Teryaev, N.S.~Tsegelnik and V.I.~Zakharov, \emph{{Modeling of acceleration in heavy-ion collisions: occurrence of temperature below the Unruh temperature}},  \href{https://arxiv.org/abs/2502.10146}{{\ttfamily 2502.10146}}.

\bibitem{Kou:2024dml}
W.~Kou and X.~Chen, \emph{{Locating quark-antiquark string breaking in QCD through chiral symmetry restoration and Hawking-Unruh effect}}, \href{https://doi.org/10.1016/j.physletb.2024.138942}{\emph{Phys. Lett. B} {\bfseries 856} (2024) 138942} [\href{https://arxiv.org/abs/2405.18697}{{\ttfamily 2405.18697}}].

\bibitem{Tolman:1930ona}
R.~Tolman and P.~Ehrenfest, \emph{{Temperature Equilibrium in a Static Gravitational Field}}, \href{https://doi.org/10.1103/PhysRev.36.1791}{\emph{Phys. Rev.} {\bfseries 36} (1930) 1791}.

\bibitem{Chernodub:2024wis}
M.N.~Chernodub, V.A.~Goy, A.V.~Molochkov, D.V.~Stepanov and A.S.~Pochinok, \emph{{Extreme Softening of QCD Phase Transition under Weak Acceleration: First-Principles Monte~Carlo Results for Gluon Plasma}}, \href{https://doi.org/10.1103/PhysRevLett.134.111904}{\emph{Phys. Rev. Lett.} {\bfseries 134} (2025) 111904} [\href{https://arxiv.org/abs/2409.01847}{{\ttfamily 2409.01847}}].

\bibitem{Crispino:2007eb}
L.C.B.~Crispino, A.~Higuchi and G.E.A.~Matsas, \emph{{The Unruh effect and its applications}}, \href{https://doi.org/10.1103/RevModPhys.80.787}{\emph{Rev. Mod. Phys.} {\bfseries 80} (2008) 787} [\href{https://arxiv.org/abs/0710.5373}{{\ttfamily 0710.5373}}].

\bibitem{Braguta:2026nfy}
V.~Braguta, V.~Goy, J.~Dey and A.~Roenko, \emph{{Spatial confinement-deconfinement transition in accelerated gluodynamics within lattice simulation}},  \href{https://arxiv.org/abs/2602.20970}{{\ttfamily 2602.20970}}.

\bibitem{Curci:1983an}
G.~Curci, P.~Menotti and G.~Paffuti, \emph{{Symanzik's Improved Lagrangian for Lattice Gauge Theory}}, \href{https://doi.org/10.1016/0370-2693(83)91043-2}{\emph{Phys. Lett. B} {\bfseries 130} (1983) 205}.

\bibitem{Luscher:1985zq}
M.~Luscher and P.~Weisz, \emph{{Computation of the Action for On-Shell Improved Lattice Gauge Theories at Weak Coupling}}, \href{https://doi.org/10.1016/0370-2693(85)90966-9}{\emph{Phys. Lett. B} {\bfseries 158} (1985) 250}.

\bibitem{Beinlich:1997ia}
B.~Beinlich, F.~Karsch, E.~Laermann and A.~Peikert, \emph{{String tension and thermodynamics with tree level and tadpole improved actions}}, \href{https://doi.org/10.1007/s100520050326}{\emph{Eur. Phys. J. C} {\bfseries 6} (1999) 133} [\href{https://arxiv.org/abs/hep-lat/9707023}{{\ttfamily hep-lat/9707023}}].

\bibitem{Borsanyi:2022xml}
S.~Borsanyi, R.~Kara, Z.~Fodor, D.A.~Godzieba, P.~Parotto and D.~Sexty, \emph{{Precision study of the continuum SU(3) Yang-Mills theory: How to use parallel tempering to improve on supercritical slowing down for first order phase transitions}}, \href{https://doi.org/10.1103/PhysRevD.105.074513}{\emph{Phys. Rev. D} {\bfseries 105} (2022) 074513} [\href{https://arxiv.org/abs/2202.05234}{{\ttfamily 2202.05234}}].

\bibitem{Kaczmarek:2002mc}
O.~Kaczmarek, F.~Karsch, P.~Petreczky and F.~Zantow, \emph{{Heavy quark anti-quark free energy and the renormalized Polyakov loop}}, \href{https://doi.org/10.1016/S0370-2693(02)02415-2}{\emph{Phys. Lett. B} {\bfseries 543} (2002) 41} [\href{https://arxiv.org/abs/hep-lat/0207002}{{\ttfamily hep-lat/0207002}}].

\bibitem{Gupta:2007ax}
S.~Gupta, K.~Huebner and O.~Kaczmarek, \emph{{Renormalized Polyakov loops in many representations}}, \href{https://doi.org/10.1103/PhysRevD.77.034503}{\emph{Phys. Rev. D} {\bfseries 77} (2008) 034503} [\href{https://arxiv.org/abs/0711.2251}{{\ttfamily 0711.2251}}].

\bibitem{Braguta:2021jgn}
V.V.~Braguta, A.Y.~Kotov, D.D.~Kuznedelev and A.A.~Roenko, \emph{{Influence of relativistic rotation on the confinement-deconfinement transition in gluodynamics}}, \href{https://doi.org/10.1103/PhysRevD.103.094515}{\emph{Phys. Rev. D} {\bfseries 103} (2021) 094515} [\href{https://arxiv.org/abs/2102.05084}{{\ttfamily 2102.05084}}].

\bibitem{Braguta:2023iyx}
V.V.~Braguta, M.N.~Chernodub and A.A.~Roenko, \emph{{New mixed inhomogeneous phase in vortical gluon plasma: First-principle results from rotating SU(3) lattice gauge theory}}, \href{https://doi.org/10.1016/j.physletb.2024.138783}{\emph{Phys. Lett. B} {\bfseries 855} (2024) 138783} [\href{https://arxiv.org/abs/2312.13994}{{\ttfamily 2312.13994}}].

\bibitem{Braguta:2024zpi}
V.V.~Braguta, M.N.~Chernodub, Y.A.~Gershtein and A.A.~Roenko, \emph{{On the origin of mixed inhomogeneous phase in vortical gluon plasma}}, \href{https://doi.org/10.1007/JHEP09(2025)079}{\emph{JHEP} {\bfseries 09} (2025) 079} [\href{https://arxiv.org/abs/2411.15085}{{\ttfamily 2411.15085}}].

\end{thebibliography}\endgroup

\end{document}